\documentclass[11pt]{article}

\usepackage {acl}
\usepackage{microtype}
\usepackage{amsmath}
\usepackage{amssymb}
\usepackage{booktabs}
\usepackage{makecell}
\usepackage[english]{babel}
\usepackage[T1]{fontenc}
\usepackage[utf8]{inputenc}
\usepackage{times}
\usepackage{booktabs}
\usepackage{multirow}
\usepackage{tabularx}
\usepackage{array}
\usepackage{ragged2e}
\usepackage[table]{xcolor}
\usepackage{booktabs}
\usepackage{enumitem}
\usepackage{xcolor}
\newcommand{\gt}[1]{\textcolor{red}{#1}}
\newcommand{\keyev}[1]{\textcolor{blue}{#1}}

\usepackage[table]{xcolor}
\definecolor{motivered}{RGB}{192,0,0}
\definecolor{motivegreen}{RGB}{0,128,0}
\usepackage{graphicx}
\title{DyHist: Query-Aware Historical Dynamics Modeling for Temporal Knowledge Graph Extrapolation}

\author{
 \textbf{Peiyuan Jiang\textsuperscript{1}},
 \textbf{Yao Liu\textsuperscript{1,*}},
 \textbf{Yubo Wen\textsuperscript{1}},
 \textbf{Yanglei Gan\textsuperscript{2}},
 \textbf{Lu Liu\textsuperscript{1}},
 \textbf{Ruojie Zhang\textsuperscript{1}}
\\
 \textbf{Jiaye Yang\textsuperscript{1}},
 \textbf{Yuewei Xue\textsuperscript{1}},
 \textbf{Yajiao Deng\textsuperscript{1}},
 \textbf{Xiaoyu Guo\textsuperscript{1}},
 \textbf{Qiao Liu\textsuperscript{1}}
\\
 \textsuperscript{1}University of Electronic Science and Technology of China, Chengdu, China\\
 \textsuperscript{2}Southwest Minzu University, Chengdu, China
\\
 \small{
   \textsuperscript{*}\textbf{Correspondence:}
   \href{mailto:liuyao@uestc.edu.cn}{liuyao@uestc.edu.cn}
 }
}
\begin{document}

\maketitle

\begin{abstract}
Temporal Knowledge Graph (TKG) extrapolation aims to predict future facts from observed historical events.
Existing methods usually achieve this goal by jointly modeling local historical evolution and long-range historical contexts.
First, local evolution modeling often summarizes local history into an overall evolving representation, where query-relevant local deviations may not be sufficiently exploited as useful predictive features.
Second, existing long-range history modeling often treats historical recurrence as a stable cue, while overlooking that its reliability may vary across relations and recent contexts.
To address these issues, we propose \textbf{DyHist}, a query-aware historical dynamics modeling framework for TKG extrapolation.
DyHist captures local dynamics by separating recent evidence from medium-range context and explicitly modeling their query-relevant deviations as predictive features.
It further uses a relation-aware historical prior to calibrate candidate scores with time-decayed support, recent trends, and relation-specific strength, enabling recurrence signals to contribute differently across relations and recent contexts.
Experiments on four benchmarks show that DyHist achieves the best MRR against recent baselines.
Code is available at \url{https://github.com/PANPANKK/DyHist}.
\end{abstract}

\begin{figure*}[t]
    \centering
    \includegraphics[width=1.0\textwidth]{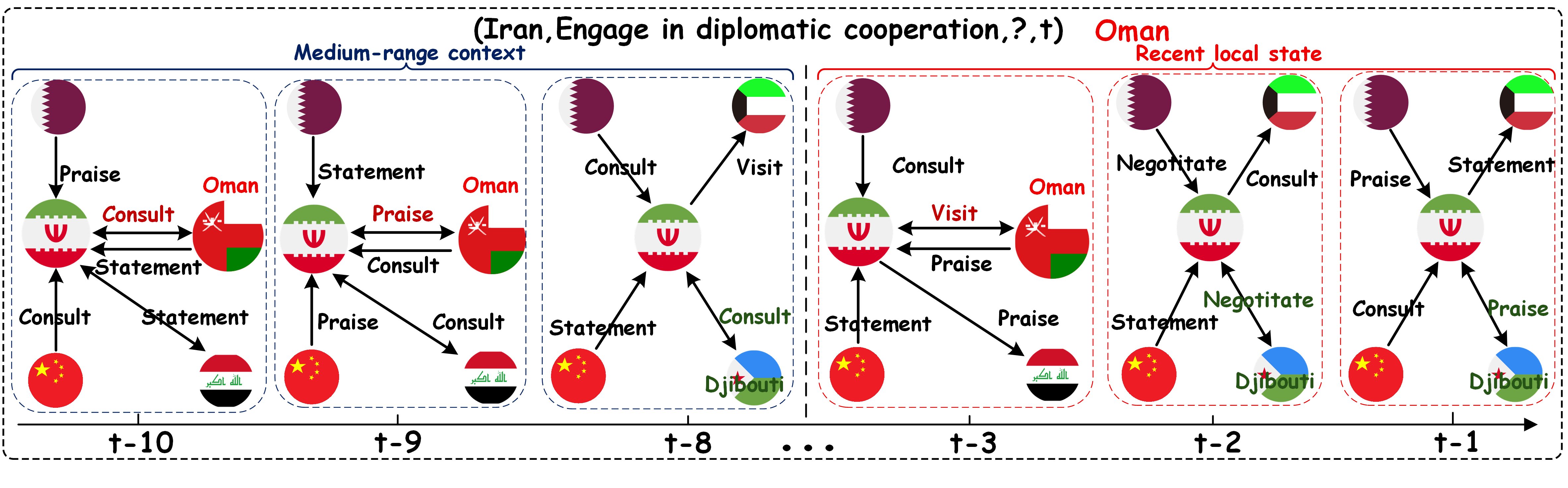}
\caption{
Illustration of local historical evidence for the query
$(\mathrm{Iran}, \mathrm{Engage\ in\ diplomatic\ cooperation}, ?, t)$.
\textcolor{motivered}{Red links} denote \textcolor{motivered}{Oman} as medium-range contextual evidence, while
\textcolor{motivegreen}{green links} denote \textcolor{motivegreen}{Djibouti} as more recent evidence near the query time.
}
    \label{fig:motivation}
\end{figure*}

\section{Introduction}
Knowledge Graphs (KGs) represent facts as triples $(s,r,o)$, but many real-world facts evolve over time~\cite{sun2020colake,wang2021covid,xu2023temporal}. 
Temporal Knowledge Graphs (TKGs) therefore extend triples into timestamped facts $(s,r,o,t)$, enabling reasoning over time-dependent facts~\cite{zhu2021learning}. 
TKG reasoning includes interpolation over observed timestamps and extrapolation to future timestamps. 
This paper focuses on extrapolation, which is important for early warning and decision-making~\cite{Li21Search}.

Accurate TKG extrapolation relies on deeply exploiting historical facts.
Given a future query $(s,r,?,t)$, the model needs to infer the missing object by exploiting historical dependencies among entities and relations. 
Existing methods have improved this task by modeling both short-term local evolution and long-range historical recurrence~\cite{li2022tirgn,chen2024local,tang2025anre}.
For short-term local history modeling, existing methods typically encode adjacent historical snapshots with GNN-based spatial encoders and recurrent or attention-based temporal modules to capture short-range structural evolution.
Beyond adjacent-snapshot encoding, recent methods further enhance local history modeling with query-related paths, subgraphs, neighborhoods, or high-order structural associations~\cite{han2020xerte,chen2025cogntke,zhang2023learning,ahrabian2026toward}.
For long-range history modeling, existing methods exploit distant historical facts mainly from two perspectives.
Some methods extract explicit recurrence signals from historical candidates, occurrence records, or frequency statistics, while others encode query-related global contexts to capture implicit long-range dependencies~\cite{zhu2021learning,xu2023temporal,ye2026dual}.
Despite these advances, existing methods still have two limitations in historical information utilization.
\textbf{First}, local history is usually encoded into a unified evolving representation through graph-based message passing and recurrent temporal aggregation.
Although this captures the overall temporal evolution, the difference between earlier context and recent evidence is usually only implicitly fused into the evolving state, rather than being explicitly modeled as a predictive signal.
As illustrated in Figure~\ref{fig:motivation}, Oman mainly appears in earlier local snapshots, whereas Djibouti becomes more prominent in recent snapshots.
Compressing the whole local sequence into a unified representation may obscure such query-relevant shifts between earlier context and recent evidence.
\textbf{Second}, existing long-range history modeling often exploits repeated co-occurrence as recurrence evidence, while underexploring how its reliability may vary across relations and recent contexts.
For relatively stable relations such as \textit{Cooperate}, recurrence may provide useful cues due to persistent diplomatic or economic ties.
For more event-driven relations such as \textit{Accuse} or \textit{Make statement}, however, candidate entities may shift with recent events, making recurrence less consistently reliable.

To address these issues, we propose \textbf{DyHist}, a query-aware framework for modeling historical dynamics in TKG extrapolation.
DyHist models historical dynamics from both local and long-range perspectives.
For local history, DyHist uses query-aware snapshot descriptors to adaptively select non-overlapping recent evidence and medium-range context.
The selected evidence is fused into a broader local historical context, which is then compared with the latest short-term evidence.
DyHist explicitly models their deviation through fast--slow residual injection with a learned gate, allowing query-relevant recent changes to complement the overall local representation.
For long-range history, DyHist introduces a relation-aware historical prior to calibrate candidate scores.
For each candidate entity, the prior aggregates previous occurrences of the complete fact $(s,r,o)$ with temporal decay, measures whether its historical support has recently increased, and scales the resulting recurrence evidence with a learnable relation-specific strength.
The main contributions of this paper are as follows:
\begin{itemize}[leftmargin=*]

\item We propose DyHist, a query-aware historical dynamics modeling framework for TKG extrapolation that jointly models recent evidence, medium-range context, and long-range historical recurrence for more accurate future fact prediction.

\item We introduce query-aware dual-window selection and fast--slow residual injection for local history modeling. The model adaptively selects non-overlapping query-relevant recent and medium-range evidence and explicitly captures the deviation between the latest short-term evidence and the broader local historical context.

\item We introduce a relation-aware dynamic historical prior that calibrates candidate scores using time-decayed complete-fact recurrence, recent support trends, and learnable relation-specific strength. Extensive experiments on four datasets demonstrate the effectiveness of DyHist.

\end{itemize}

\begin{figure*}[t]
    \centering
    \includegraphics[width=1.0\textwidth]{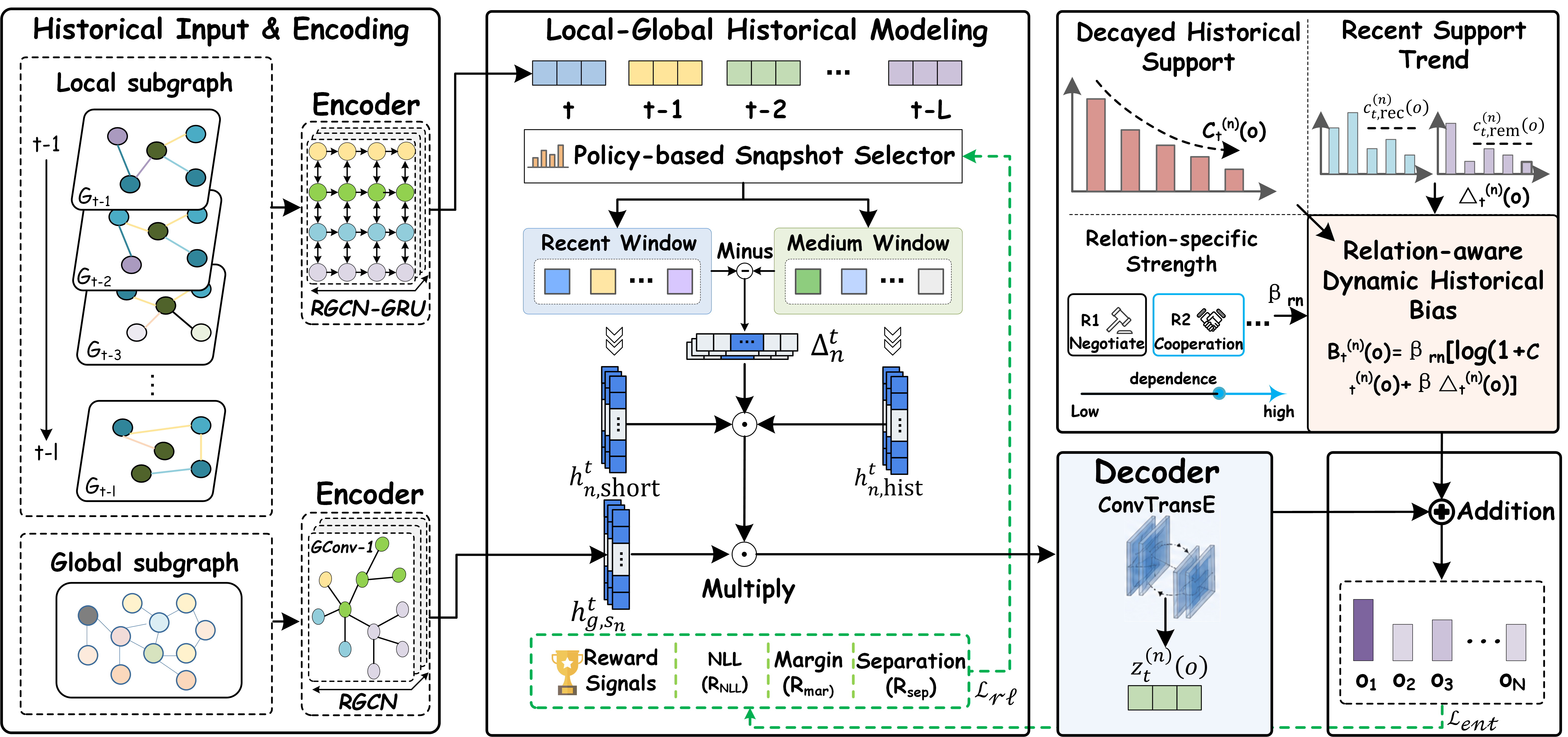}
\caption{Overall architecture of DyHist.
DyHist models query-aware local dynamics through dual-window selection and fast--slow residual injection, and calibrates decoder logits with a relation-aware dynamic historical prior.}
    \label{fig:framework}
\end{figure*}

\section{Related Work}
\subsection{TKG Extrapolation}
Temporal Knowledge Graph (TKG) extrapolation aims to predict future facts from observed historical facts.
A common paradigm is dynamic representation learning, which encodes historical events or temporal snapshots to learn evolving entity and relation representations.
Representative methods include Know-Evolve, which models temporal facts with point processes~\citep{trivedi2017know}, RE-NET, which captures historical event sequences and concurrent structures~\citep{jin2020recurrent}, and RE-GCN, which combines graph neural networks with recurrent evolution over snapshots~\citep{li2021temporal}.

Recent studies further improve TKG extrapolation from diverse perspectives. TLogic performs interpretable reasoning using temporal logical rules~\citep{liu2022tlogic}. DiffuTKG and NADEx model future entities or events from a generative perspective~\citep{cai2024predicting,gan26negative}. HTKGE focuses on temporal scoring and decoder design~\citep{yang2024unleashing}, while AnRe incorporates textual semantics and textualized historical contexts for temporal reasoning~\citep{tang2025anre}.
Although these methods differ in modeling paradigms, organizing and exploiting historical information remains central to TKG extrapolation.

\subsection{Historical Information Modeling}
Historical information modeling is a key component of TKG extrapolation.
Existing approaches can be broadly grouped into local-history modeling and long-range history modeling. Local-history modeling focuses on facts close to the query timestamp, aiming to capture recent entity states and short-term structural evolution.
Existing strategies include adjacent-snapshot encoding, temporal-neighborhood aggregation, query-related path reasoning, local-subgraph construction, and causal-subhistory modeling.
Representative methods include RE-NET, RE-GCN, xERTE, TITer, CSI, and CognTKE~\citep{jin2020recurrent,li2021temporal,han2020xerte,sun2021timetraveler,chen2024temporal,chen2025cogntke}.
These methods provide informative local evidence for future prediction.
However, recent signals and earlier contextual information are often aggregated into a unified representation, making query-relevant local changes difficult to model explicitly. Long-range history modeling exploits distant historical facts to capture repeated patterns and long-range dependencies.
One line introduces explicit recurrence signals, such as repeated facts, historical candidates, occurrence records, or frequency statistics~\citep{zhu2021learning,xu2023temporal,li2022tirgn,wang2026temporal}.
Another line encodes global historical facts, query-related subgraphs, or retrieved long-range contexts into latent representations for reasoning~\citep{chen2024local,ye2026dual,chen2025cogntke,tang2025anre}.
These studies show that long-range history provides complementary evidence beyond recent snapshots.
Nevertheless, recurrence evidence is often incorporated without explicitly calibrating its reliability across relations and recent temporal contexts.

In contrast to existing approaches, DyHist explicitly distinguishes recent evidence, medium-range local context, and long-range historical recurrence.
It uses query-aware dual-window selection and fast--slow residual injection to model query-relevant local changes, while a relation-aware historical prior calibrates the influence of recurring historical patterns across relations and recent contexts.

\section{Method}
\label{sec:method}

\subsection{Overview}
\label{subsec:overview}
As shown in Figure~\ref{fig:framework}, DyHist targets TKG extrapolation, which predicts the missing object in a future query $(s_n,r_n,?,t)$ from snapshots before $t$.
Given $\{\mathcal{G}_{1},\ldots,\mathcal{G}_{t-1}\}$, DyHist uses a local sequence
$\mathcal{H}_t^{\mathrm{loc}}=\{\mathcal{G}_{t-L},\ldots,\mathcal{G}_{t-1}\}$
to model recent evolution and a sampled global subgraph $\mathcal{G}_t^{\mathrm{glob}}$ to capture long-range evidence.
The local sequence and global subgraph are encoded by RGCN-GRU and RGCN encoders, respectively.
Let $\mathbf{e}_{s_n}^{0}$ and $\mathbf{e}_{r_n}^{0}$ denote the learnable base embeddings of subject entity $s_n$ and relation $r_n$.
DyHist constructs the query representation as
$\mathbf{q}_n=W_q[\mathbf{e}_{s_n}^{0}\Vert\mathbf{e}_{r_n}^{0}]$,
where $W_q$ is a learnable projection matrix.
Using query-aware snapshot descriptors, DyHist selects recent and medium windows from the local sequence, fuses the selected evidence into a local context, and enhances it with fast--slow residual injection.
The enhanced local representation is then combined with the global representation for scoring, while a relation-aware dynamic historical prior calibrates logits using recurrence evidence. Main notation is summarized in Appendix~\ref{app:notation}.
\subsection{Historical Representation Encoding}
\label{subsec:history_encoding}
For each local snapshot $\mathcal{G}_{\tau}$, DyHist updates local entity states by combining RGCN-based structural encoding with recurrent transition:
\begin{equation}
\mathbf{H}_{l}^{\tau}
=
\mathrm{GRU}
\big(
\mathrm{RGCN}_{\mathrm{loc}}(\mathcal{G}_{\tau}, \mathbf{H}_{l}^{\tau-1}),
\mathbf{H}_{l}^{\tau-1}
\big).
\end{equation}
Here, $\mathbf{H}_{l}^{\tau}\in\mathbb{R}^{|\mathcal{E}|\times d}$ denotes the local entity-state matrix at timestamp $\tau$.
In addition, DyHist encodes the sampled global subgraph to obtain a global historical representation:
\begin{equation}
\mathbf{H}_{g}^{t}
=
\mathrm{RGCN}_{\mathrm{glob}}
(\mathcal{G}_t^{\mathrm{glob}}),
\end{equation}
where $\mathbf{H}_{g}^{t}$ denotes the query-time global historical representation for decoding.

\subsection{Query-aware Local Dynamics Modeling}
\label{subsec:dual_window}

\paragraph{Snapshot descriptor construction.}
For each candidate snapshot $\mathcal{G}_{\tau}$, DyHist constructs a compact descriptor
$\mathbf{d}_{\tau}=[u_{\tau},\ell_{\tau},\delta_{\tau},\alpha_{\tau},m_{\tau},m_{\tau}(1-\ell_{\tau})]$,
where $u_{\tau}$, $\ell_{\tau}$, $\delta_{\tau}$, $\alpha_{\tau}$, and $m_{\tau}$ are scalar features characterizing query relevance, temporal lag, snapshot variation, latest-state consistency, and relation-context relevance, respectively.
The last term highlights recent relation-aligned snapshots.
Detailed definitions are provided in Appendix~\ref{app:descriptor_details}.

\paragraph{Policy-based dual-window selection.}
Let $\mathcal{I}_t^{\mathrm{loc}}=\{t-L,\ldots,t-1\}$ denote the local candidate snapshot indices.
Based on the descriptor set
$\mathcal{D}_t=\{\mathbf{d}_{\tau}\mid \tau\in\mathcal{I}_t^{\mathrm{loc}}\}$,
DyHist samples recent and medium window index sets:
\begin{equation}
\mathcal{I}_t^{R}\sim\pi_R(\cdot\mid\mathcal{D}_t),
\qquad
\mathcal{I}_t^{M}\sim\pi_M(\cdot\mid\mathcal{D}_t),
\end{equation}
where $\pi_R$ and $\pi_M$ are role-specific policies over candidate snapshot indices.
The recent window $\mathcal{I}_t^{R}$ is selected first, after which the medium
window $\mathcal{I}_t^{M}$ is selected from
$\mathcal{I}_t^{\mathrm{loc}}\setminus\mathcal{I}_t^{R}$,
ensuring non-overlapping temporal evidence.
The recent window captures near-query dynamics, while the medium window provides earlier contextual evidence.

\paragraph{Dual-window aggregation.}
Given the query representation $\mathbf{q}_n$, DyHist separately aggregates the encoded subject states in each selected window:
\begin{equation}
\mathbf{h}_{n}^{X}
=
\mathrm{Agg}_{X}
(\{\mathbf{H}_{l}^{\tau}[s_n]\}_{\tau\in\mathcal{I}_t^{X}};\mathbf{q}_n).
\end{equation}
Here, $X$ denotes either the recent window $R$ or the medium window $M$, and $\mathrm{Agg}_{X}$ is a query-guided temporal aggregator.
Its implementation is detailed in Appendix~\ref{app:aggregation_details}. The two window representations are fused through a vector gate:
\begin{equation}
\begin{aligned}
\mathbf{g}_{n}
&=
\sigma
\big(W_g[\mathbf{h}_{n}^{R}\|\mathbf{h}_{n}^{M}]\big), \\
\mathbf{h}_{n}^{L}
&=
\mathbf{g}_{n}\odot \mathbf{h}_{n}^{R}
+
(1-\mathbf{g}_{n})\odot \mathbf{h}_{n}^{M}.
\end{aligned}
\end{equation}
where $\sigma(\cdot)$ is the sigmoid function and $\mathbf{g}_{n}$ is a vector gate that balances near-query dynamic evidence and medium-range contextual evidence.
\paragraph{Fast-slow residual injection.}
To explicitly capture query-relevant local changes, DyHist compares the latest short-term evidence with the fused local historical context:
\begin{equation}
\Delta_n^t
=
\mathbf{h}_{n,\mathrm{short}}^t
-
\mathbf{h}_{n,\mathrm{hist}}^t,
\qquad
\mathbf{h}_{n,\mathrm{hist}}^t
=
\mathbf{h}_{n}^{L}.
\end{equation}
Here, $\mathbf{h}_{n,\mathrm{short}}^t$ is obtained by applying the same query-guided aggregator to the most recent subset
$\mathcal{I}_t^S\subseteq\mathcal{I}_t^R$, with details in Appendix~\ref{app:aggregation_details}.
The residual $\Delta_n^t$ characterizes the deviation between near-query evidence and the broader local historical context.
The final local subject representation is obtained by gated residual injection:
\begin{equation}
\mathbf{h}_{l,s_n}^{t}
=
\mathbf{h}_{n,\mathrm{hist}}^{t}
+
\sigma_g(\mathbf{h}_{n,\mathrm{hist}}^{t},\Delta_n^t)
\odot
\Delta_n^t ,
\end{equation}
where $\sigma_g(\cdot)$ denotes a sigmoid-based vector gate.

\subsection{Dynamic Historical Prior}
\label{subsec:decoder}
For query $(s_n,r_n,?,t)$, DyHist first combines local and global historical representations for decoding:
\begin{equation}
\mathbf{h}_{s_n,t}^{\mathrm{dec}}
=
\lambda \mathbf{h}_{l,s_n}^{t}
+
(1-\lambda)\mathbf{H}_{g}^{t}[s_n],
\end{equation}
where $\lambda$ balances local and global historical information.
The decoder then produces neural logits over all candidate objects using the context-aware relation representation $\mathbf{e}_{r_n,t}$:
\begin{equation}
\mathbf{x}_{n,t}
=
f_{\mathrm{dec}}(\mathbf{h}_{s_n,t}^{\mathrm{dec}},\mathbf{e}_{r_n,t}),
\mathbf{z}_{t}^{(n)}
=
\mathbf{x}_{n,t}(\mathbf{H}_{l}^{t-1})^{\top}.
\end{equation}
Here, $\mathbf{H}_{l}^{t-1}$ serves as the candidate object representation bank, and $\mathbf{z}_{t}^{(n)}[o]$ denotes the neural logit of candidate object $o$.
To calibrate candidate logits with recurrence evidence, DyHist first computes time-decayed support for each candidate:
\begin{equation}
c_t^{(n)}(o)
=
\sum_{\tau<t}
e^{-\gamma(t-\tau)}
\mathbb{I}[(s_n,r_n,o)\in\mathcal{G}_{\tau}].
\end{equation}
Let $\mathcal{I}_t^{\mathrm{rem}}=\mathcal{I}_t^{\mathrm{loc}}\setminus\mathcal{I}_t^{R}$ denote the remaining local indices.
Recent support changes are captured by
\begin{equation}
\eta_t^{(n)}(o)
=
\mathrm{ReLU}
\left(
\frac{c_{t,\mathrm{rec}}^{(n)}(o)}{|\mathcal{I}_t^{R}|}
-
\frac{c_{t,\mathrm{rem}}^{(n)}(o)}
{|\mathcal{I}_t^{\mathrm{rem}}|}
\right),
\end{equation}
where $c_{t,\mathrm{rec}}^{(n)}(o)$ and $c_{t,\mathrm{rem}}^{(n)}(o)$ count support in $\mathcal{I}_t^R$ and $\mathcal{I}_t^{\mathrm{rem}}$, respectively.
A learnable relation-specific strength $\beta_r$ controls the prior reliability.
The relation-aware historical bias is computed as
\begin{equation}
b_t^{(n)}(o)
=
\beta_{r_n}
\left[
\log(1+c_t^{(n)}(o))
+
\rho\eta_t^{(n)}(o)
\right],
\end{equation}
where $\beta_{r_n}$ controls the recurrence strength of relation $r_n$ and $\rho$ controls the trend contribution.
Finally, the calibrated logit is
\begin{equation}
\hat{z}_t^{(n)}(o)
=
z_t^{(n)}(o)
+
b_t^{(n)}(o).
\end{equation}

\subsection{Training Objective}
\label{subsec:objective}

\paragraph{Entity prediction loss.}
For the target timestamp $t$ with $N_t$ facts
$\{(s_n,r_n,o_n,t)\}_{n=1}^{N_t}$,
DyHist minimizes the object prediction negative log-likelihood:
\begin{equation}
\mathcal{L}_{\mathrm{ent}}
=
-\frac{1}{N_t}
\sum_{n=1}^{N_t}
\log p(o_n\mid s_n,r_n,t).
\end{equation}

\paragraph{Window selection loss.}
The dual-window policy is optimized with prediction-oriented reward feedback and temporal-role separation:
\begin{equation}
\mathcal{R}_t
=
\lambda_m\mathcal{R}_{\mathrm{mar}}
+
\lambda_n\mathcal{R}_{\mathrm{nll}}
+
\mathcal{R}_{\mathrm{sep}},
\end{equation}
where $\mathcal{R}_{\mathrm{mar}}$ measures the prediction margin,
$\mathcal{R}_{\mathrm{nll}}=-\mathcal{L}_{\mathrm{ent}}$ provides loss-based feedback,
and $\mathcal{R}_{\mathrm{sep}}$ encourages distinct temporal roles.
Given the joint window-selection action $a_t$ and descriptor set $\mathcal{D}_t$, the policy-gradient loss is
\begin{equation}
\mathcal{L}_{\mathrm{pg}}
=
-
(\mathcal{R}_t-\bar{b}_t)
\log \pi(a_t\mid \mathcal{D}_t)
-
\lambda_{\mathrm{H}}\mathcal{H}(\pi),
\end{equation}
where $\bar{b}_t$ is the reward baseline and $\mathcal{H}(\pi)$ is the entropy regularizer.
Detailed reward definitions are given in Appendix~\ref{app:pg_objective}.

\paragraph{Overall objective.}
The final objective is
\begin{equation}
\mathcal{L}
=
\mathcal{L}_{\mathrm{ent}}
+
\lambda_{\mathrm{pg}}
\mathcal{L}_{\mathrm{pg}},
\end{equation}
where $\lambda_{\mathrm{pg}}$ balances prediction and policy losses.

\definecolor{bestgray}{gray}{0.78}
\definecolor{secondgray}{gray}{0.90}
\newcommand{\best}[1]{\cellcolor{bestgray}\textbf{#1}}
\newcommand{\second}[1]{\cellcolor{secondgray}\underline{#1}}
\begin{table*}[t]
\centering
\caption{Overall performance (\%) on four datasets in terms of MRR and Hits@1/3/10 under the time-aware filtered setting. 
Gray cells indicate the best results, and light gray cells indicate the second-best.}
\label{tab:main_results}
\footnotesize
\setlength{\tabcolsep}{2pt}
\renewcommand{\arraystretch}{0.82}
\resizebox{\textwidth}{!}{
\begin{tabular}{lcccccccccccccccc}
\toprule
\multirow{2}{*}{Models}
& \multicolumn{4}{c}{ICEWS14}
& \multicolumn{4}{c}{ICEWS18}
& \multicolumn{4}{c}{ICEWS05-15}
& \multicolumn{4}{c}{GDELT} \\
\cmidrule(lr){2-5}
\cmidrule(lr){6-9}
\cmidrule(lr){10-13}
\cmidrule(lr){14-17}
& MRR & H@1 & H@3 & H@10
& MRR & H@1 & H@3 & H@10
& MRR & H@1 & H@3 & H@10
& MRR & H@1 & H@3 & H@10 \\
\midrule
DistMult & 15.44 & 10.91 & 17.24 & 23.92 & 11.51 & 7.03 & 12.87 & 20.86 & 17.95 & 13.12 & 20.71 & 29.32 & 8.68 & 5.58 & 9.96 & 17.13 \\
ConvE & 35.09 & 25.23 & 39.38 & 54.68 & 24.51 & 16.23 & 29.25 & 44.51 & 33.81 & 24.78 & 39.00 & 54.95 & 16.55 & 11.02 & 18.88 & 31.60 \\
RotatE & 21.31 & 10.26 & 24.35 & 44.75 & 12.78 & 4.01 & 14.89 & 31.91 & 24.71 & 13.22 & 29.04 & 48.16 & 13.45 & 6.95 & 14.09 & 25.99 \\
\midrule
TTransE & 13.72 & 2.98 & 17.70 & 35.74 & 8.31 & 1.92 & 8.56 & 21.89 & 15.57 & 4.80 & 19.24 & 38.29 & 5.50 & 0.47 & 4.94 & 15.25 \\
TA-DistMult & 25.80 & 16.94 & 29.74 & 42.99 & 16.75 & 8.61 & 18.41 & 33.59 & 24.31 & 14.58 & 27.92 & 44.21 & 12.00 & 5.76 & 12.94 & 23.54 \\
DE-SimplE & 33.36 & 24.85 & 37.15 & 48.92 & 19.30 & 11.53 & 21.86 & 34.80 & 35.02 & 25.91 & 38.99 & 52.75 & 19.70 & 12.22 & 21.39 & 33.70 \\
\midrule
RE-NET & 36.93 & 26.83 & 39.51 & 54.78 & 28.81 & 19.05 & 32.44 & 47.51 & 43.32 & 33.43 & 47.77 & 63.06 & 19.62 & 12.42 & 21.00 & 34.01 \\
RE-GCN & 40.39 & 30.66 & 44.96 & 59.21 & 30.58 & 21.01 & 34.34 & 48.75 & 48.03 & 37.33 & 53.85 & 68.27 & 19.64 & 12.42 & 20.90 & 33.69 \\
CyGNet & 35.05 & 25.73 & 39.01 & 53.55 & 24.93 & 15.90 & 28.28 & 42.61 & 36.81 & 26.61 & 41.63 & 56.22 & 18.48 & 11.52 & 19.57 & 31.98 \\
TiRGN & 44.04 & 33.83 & 48.95 & 63.84 & 33.66 & 23.19 & 37.99 & 54.22 & 50.04 & 39.25 & 56.13 & 70.71 & 21.67 & 13.63 & 23.27 & 37.60 \\
HisMatch & 46.42 & 35.91 & 51.63 & 66.84 & 33.99 & 23.91 & 37.90 & 53.94 & 52.85 & 42.01 & 59.05 & 73.28 & 22.01 & 14.45 & 23.80 & 36.61 \\
RETIA & 42.76 & 32.28 & 47.77 & 62.75 & 32.43 & 22.23 & 36.48 & 52.94 & 47.26 & 36.64 & 52.90 & 67.76 & 20.12 & 12.76 & 21.45 & 34.49 \\
CENET & 39.02 & 29.62 & 43.23 & 57.49 & 27.85 & 18.15 & 31.63 & 46.98 & 41.95 & 32.17 & 46.93 & 60.43 & 20.23 & 12.69 & 21.70 & 34.92 \\
CRAFT & 45.71 & 35.05 & 51.83 & 65.21 & 34.21 & 23.96 & 38.53 & 54.11 & 50.14 & 39.56 & 56.18 & 70.09 & 23.78 & 15.38 & 26.23 & 40.15 \\
THCN & 45.39 & 36.58 & 50.84 & 66.07 & 35.63 & 24.90 & 39.26 & 56.76 & 51.94 & 40.32 & 57.79 & 72.18 & 23.46 & 15.18 & 25.21 & 39.03 \\
LogCL & 48.87 & 37.76 & 54.71 & 70.26 & 35.67 & 24.53 & 40.32 & 57.74 & 57.07 & 46.07 & 63.72 & 77.87 & 23.75 & 14.64 & 25.60 & 42.33 \\
CognTKE & 46.06 & 36.49 & 51.11 & 64.49 & 35.24 & 25.21 & 39.93 & 54.71 & 53.13 & 42.62 & 59.42 & 72.70 & -- & -- & -- & -- \\
NADEx & 49.03 & 39.45 & 57.48 & 70.55 & 36.84 & \second{27.58} & \second{45.12} & \second{60.58} & 52.17 & 43.38 & 60.75 & 71.93 & 23.67 & 16.10 & \best{28.35} & 43.05 \\
MIDFA & \second{51.74} & \second{40.93} & \second{57.73} & \second{72.40} & \second{37.80} & 26.63 & 42.68 & 59.88 & \second{58.55} & \second{47.51} & \second{65.56} & \second{79.08} & \second{25.18} & \best{16.69} & 27.89 & \second{44.53} \\
\midrule
Ours & \best{54.03} & \best{43.25} & \best{60.01} & \best{74.49} & \best{40.30} & \best{28.85} & \best{45.59} & \best{62.59} & \best{61.07} & \best{50.46} & \best{67.67} & \best{81.10} & \best{25.53} & \second{16.16} & \second{28.08} & \best{45.19} \\
\bottomrule
\end{tabular}
}
\end{table*}
\section{Experiments}
\label{sec:experiments}
\subsection{Experimental Setups}
\label{subsec:experimental_setups}
\paragraph{Datasets and Evaluation Metrics.}
We evaluate DyHist on four standard TKG extrapolation benchmarks: ICEWS14, ICEWS05-15, ICEWS18, and GDELT.
The ICEWS datasets~\cite{boschee2015icews} are derived from the Integrated Crisis Early Warning System, and GDELT~\cite{leetaru2013gdelt} is collected from the Global Database of Events, Language, and Tone.
Following previous studies~\cite{chen2024local}, we use chronological splits and report MRR and Hits@$\{1,3,10\}$ under the time-aware filtered setting.
Dataset statistics are provided in Appendix~\ref{app:dataset_statistics}.

\paragraph{Baseline Models.}
We compare DyHist with representative methods from three categories.
Static methods include DistMult \citep{yang2015embedding}, ConvE \citep{dettmers2018convolutional}, and RotatE \citep{sun2018rotate}.
Interpolation methods include TTransE \citep{garcia2018learning}, TA-DistMult \citep{garcia2018learning}, and DE-SimplE \citep{goel2020diachronic}.
Extrapolation methods include RE-NET~\citep{jin2020recurrent},
RE-GCN~\citep{li2021temporal},
CyGNet~\citep{zhu2021learning},
TiRGN~\citep{li2022tirgn},
HisMatch~\citep{li2022hismatch},
RETIA~\citep{liu2023retia},
CENET~\citep{xu2023temporal},
CRAFT~\citep{zhang2024modeling},
THCN~\citep{chen2024thcn},
LogCL~\citep{chen2024local},
CognTKE~\citep{chen2025cogntke},
NADEx~\citep{gan26negative},
and MIDFA~\citep{wang2026temporal}.
Baseline results are taken from their original papers under the same evaluation setting whenever available.
Baseline details are given in Appendix~\ref{app:baseline_descriptions}.

\paragraph{Implementation Details.}
We train DyHist with Adam for up to 80 epochs and select the checkpoint with the best validation MRR for testing.
The entity and relation embedding dimensions are both set to 200, and the dropout rate is 0.2.
We use an RGCN-based temporal graph encoder and ConvTransE decoder.
The local history length is 10, with both recent and medium-range window sizes set to 3, and the fast window size set to 2.
The relation-aware historical prior is applied at the logit level, and the policy loss weight is set to 0.01.
Training is conducted at the timestamp level, and DyHist experiments are conducted on a single RTX 4090 GPU.
Additional implementation details are provided in Appendix~\ref{app:experimental_details}.

\subsection{Main Results}
\label{subsec:main_results}

Table~\ref{tab:main_results} reports overall performance on four benchmark datasets. We highlight three observations.
First, temporal extrapolation methods generally outperform static and interpolation-based baselines, showing the importance of historical modeling for future fact prediction.
Second, DyHist achieves the highest MRR on all four datasets and obtains competitive Hits@K results.
Third, compared with recent extrapolation baselines such as LogCL, NADEx, and MIDFA, DyHist shows consistent MRR gains across datasets.
These results suggest that distinguishing different types of historical evidence is beneficial.
Unlike baselines that often aggregate history uniformly, DyHist separates query-relevant recent signals from medium-range context and calibrates candidate scores with a relation-aware historical prior. This enables adaptive evidence selection and history-aware score adjustment for TKG extrapolation.
On the larger and noisier GDELT dataset, DyHist still achieves the best MRR and Hits@10, indicating its effectiveness under complex event streams.

\subsection{Ablation Study}
\label{subsec:ablation}

To examine the contribution of each component, we conduct ablation studies on ICEWS14 and ICEWS05-15.
We consider six variants:
w/o QADW removes window selection and fast-slow residual injection, and instead performs query-aware aggregation over all local snapshots in $\mathcal{H}_t^{\mathrm{loc}}$;
w/o PG removes the policy-gradient loss for window selection;
w/o FS removes the fast-slow residual injection;
w/o Prior removes the relation-aware dynamic historical prior;
w/o Decay removes time-decayed recurrence support;
and w/o Trend removes the recent support trend.

The ablation results illustrate the contributions of DyHist's components.
For local history modeling, the drop of w/o QADW suggests the overall benefit of explicit window selection and fast-slow local dynamics modeling.
The declines of w/o PG and w/o FS further indicate the contributions of reward-driven selection and fast-slow residual injection within this module.
Among the long-range variants, w/o Prior shows the largest MRR drop, while the drops of w/o Decay and w/o Trend indicate the usefulness of decayed and trend-based recurrence signals.
Overall, these findings suggest that DyHist benefits from adaptive local evidence modeling and dynamic historical prior calibration.

\begin{table}[h]
\centering
\footnotesize
\setlength{\tabcolsep}{3pt}
\caption{Component ablation results on ICEWS14 and ICEWS05-15. Bold numbers denote the best results.}
\label{tab:ablation}
\renewcommand{\arraystretch}{0.9}
\begin{tabular}{lccc ccc}
\toprule
\multirow{2}{*}{Variant} 
& \multicolumn{3}{c}{ICEWS14} 
& \multicolumn{3}{c}{ICEWS05-15} \\
\cmidrule(lr){2-4} \cmidrule(lr){5-7}
& MRR & H@1 & H@10 
& MRR & H@1 & H@10 \\
\midrule
Ours       & \textbf{54.03} & \textbf{43.25} & \textbf{74.49} & \textbf{61.07} & \textbf{50.46} & \textbf{81.10} \\
w/o QADW   & 51.76 & 40.75 & 72.88 & 58.71 & 48.35 & 79.61 \\
w/o PG     & 52.94 & 41.43 & 73.80 & 59.48 & 49.76 & 80.74 \\
w/o FS     & 52.36 & 41.22 & 73.68 & 59.14 & 49.62 & 80.50 \\
w/o Prior  & 52.09 & 41.13 & 73.27 & 59.97 & 48.91 & 80.71 \\
w/o Decay  & 53.49 & 42.57 & 74.18 & 60.84 & 50.24 & 80.66 \\
w/o Trend  & 53.79 & 42.99 & 74.25 & 60.77 & 50.25 & 80.54 \\
\bottomrule
\end{tabular}
\end{table}

\subsection{Analysis of Local History Modeling}

We evaluate the query-aware dual-window selection (QADW) against four strategies.
Latest-$K$, Random-$K$, and Top-$K$ are single-window methods, which select snapshots by recency, random sampling, and deterministic descriptor ranking, respectively.
DFW is a fixed dual-window method, which uses $\{t-1,t-2,t-3\}$ as the recent window and $\{t-4,t-5,t-6\}$ as the medium-range window, without query-aware selection.

As shown in Table~\ref{tab:window_strategy}, QADW achieves the best MRR on both datasets.
Compared with Latest-$K$ and Random-$K$, the gains indicate that simple temporal or random choices are insufficient.
Top-$K$ shows the usefulness of query-aware snapshot ranking, while DFW shows the benefit and limitation of fixed recent/medium partitions.
QADW combines query-aware selection with adaptive dual-window roles, leading to better local history modeling with only about 0.08M additional parameters and similar inference time.
\begin{table}[t]
  \centering
  \footnotesize
  \setlength{\tabcolsep}{3pt}
  \renewcommand{\arraystretch}{0.9}
   \caption{Effectiveness and efficiency of local history selection strategies. 
Time and Params denote inference time and parameter size, respectively.}
  \label{tab:window_strategy}
  \begin{tabular}{lcccccc}
  \toprule
  \multirow{2}{*}{Strategy}
  & \multicolumn{3}{c}{ICEWS14}
  & \multicolumn{3}{c}{ICEWS05-15} \\
  \cmidrule(lr){2-4}
  \cmidrule(lr){5-7}
  & MRR & Time & Params
  & MRR & Time & Params \\
  \midrule
  Latest-$K$
  & 52.16 & \textbf{10.9s} & \textbf{9.15M}
  & 59.24 & 2m39s & \textbf{10.22M} \\
  Random-$K$
  & 51.27 & 11.1s & \textbf{9.15M}
  & 58.15 & 2m40s & \textbf{10.22M} \\
  Top-$K$
  & 52.34 & 11.5s & 9.23M
  & 60.72 & 2m43s & 10.30M \\
  DFW
  & 53.35 & 11.1s & \textbf{9.15M}
  & 60.46 & \textbf{2m37s} & \textbf{10.22M} \\
  QADW
  & \textbf{54.03} & 11.2s & 9.23M
  & \textbf{61.07} & 2m41s & 10.30M \\
  \bottomrule
  \end{tabular}
\end{table}

\paragraph{Query-aware descriptor analysis.}
We further examine descriptor features in the QADW selector.
Time-only removes query relevance and relation-context alignment, keeping only temporal distance and snapshot-level dynamics.
Q-shuf. randomly shuffles query representations across timestamp batches, and R-shuf. shuffles relation information to disrupt its alignment with the current query relations.
As shown in Table~\ref{tab:query_awareness}, the full descriptor setting achieves the best results on both datasets.
The competitive but lower performance of Time-only suggests that temporal and structural cues are useful but incomplete.
The drops of Q-shuf. and R-shuf. further indicate that preserving query- and relation-specific correspondence provides additional guidance for adaptive local history selection.

\paragraph{Effect of fast-slow residual injection.}
We examine whether the deviation between near-query evidence and the dual-window historical context provides useful local variation signals.
We compare four variants.
Short-only uses only the most recent short-term representation, testing whether near-query evidence alone is sufficient.
No residual uses the fused recent--medium historical context without residual injection, testing whether the dual-window context alone is sufficient.
Add residual directly adds the fast-slow residual to the historical context, testing whether the deviation between short-term evidence and historical context is useful.
Gated residual injects the residual through an adaptive gate, testing whether such deviations should be introduced selectively.
As shown in Table~\ref{tab:residual_analysis}, Short-only performs worse than No residual, indicating that relying only on near-query evidence is insufficient.
The improvement of Add residual over No residual shows that fast-slow deviations contain local variation signals.
Gated residual achieves slightly better overall performance, suggesting a modest benefit from adaptive residual injection.

\begin{table}[h]
\centering
\footnotesize
\setlength{\tabcolsep}{3pt}
\renewcommand{\arraystretch}{0.9}
\caption{Analysis of descriptor features under the QADW selector.}
\label{tab:query_awareness}
\begin{tabular}{lcccccc}
\toprule
\multirow{2}{*}{Input}
& \multicolumn{3}{c}{ICEWS14}
& \multicolumn{3}{c}{ICEWS05-15} \\
\cmidrule(lr){2-4}
\cmidrule(lr){5-7}
& MRR & H@1 & H@10 & MRR & H@1 & H@10 \\
\midrule
Time-only & 52.87 & 41.95 & 73.71 & 60.64 & 50.02 & 80.67 \\
Q-shuf.   & 53.18 & 42.32 & 73.65 & 60.15 & 49.94 & 80.31 \\
R-shuf.   & 53.24 & 42.58 & 73.79 & 60.42 & 50.15 & 80.40 \\
Full      & \textbf{54.03} & \textbf{43.25} & \textbf{74.49} & \textbf{61.07} & \textbf{50.46} & \textbf{81.10} \\
\bottomrule
\end{tabular}
\end{table}
\vspace{-5mm}
\begin{table}[h]
\centering
\footnotesize
\setlength{\tabcolsep}{3pt}
\renewcommand{\arraystretch}{0.9}
\caption{Analysis of fast-slow residual injection.}
\label{tab:residual_analysis}
\begin{tabular}{lcccccc}
\toprule
\multirow{2}{*}{Variant}
& \multicolumn{3}{c}{ICEWS14}
& \multicolumn{3}{c}{ICEWS05-15} \\
\cmidrule(lr){2-4}
\cmidrule(lr){5-7}
& MRR & H@1 & H@10 & MRR & H@1 & H@10 \\
\midrule
Short-only     & 51.51 & 40.27 & 73.42 & 60.00 & 49.41 & 80.06 \\
No residual    & 52.36 & 40.92 & 73.68 & 60.66 & 50.01 & 80.74 \\
Add residual   & 53.91 & 42.91 & 74.66 & 60.84 & 50.24 & 80.74 \\
Gated residual & \textbf{54.03} & \textbf{43.25} & \textbf{74.49} & \textbf{61.07} & \textbf{50.46} & \textbf{81.10} \\
\bottomrule
\end{tabular}
\end{table}
\subsection{Repeated and Novel Queries}
\label{subsec:history_analysis}
We split test queries into repeated and novel subsets according to whether the ground-truth object has appeared with the same subject--relation pair in history.
Repeated queries evaluate historical recurrence, while novel queries test behavior without direct recurrence.
We compare DyHist with Latest, which selects the latest-$K$ snapshots without query-aware descriptors, and \textbf{w/o Prior}, which removes the relation-aware historical prior.
As shown in Table~\ref{tab:repeat_novel}, repeated queries perform better than novel ones, confirming recurrence as a strong signal.
DyHist improves over w/o Prior on repeated queries, showing that the dynamic prior helps calibrate recurrence patterns.
On novel queries, DyHist slightly outperforms Latest, suggesting that query-aware descriptors can identify useful local evidence.
Overall, DyHist benefits from recurrence calibration on repeated queries and query-aware local evidence on novel queries.
\begin{table}[t]
\centering
\footnotesize
\setlength{\tabcolsep}{3pt}
\renewcommand{\arraystretch}{0.9}
\caption{Results on repeated and novel query subsets. }
\label{tab:repeat_novel}
\begin{tabular}{llcccccc}
\toprule
\multirow{2}{*}{Type}
& \multirow{2}{*}{Model}
& \multicolumn{3}{c}{ICEWS14}
& \multicolumn{3}{c}{ICEWS05-15} \\
\cmidrule(lr){3-5}
\cmidrule(lr){6-8}
& & MRR & H@1 & H@10 & MRR & H@1 & H@10 \\
\midrule
\multirow{3}{*}{Rep.}
& Latest    & 73.67 & 62.88 & \textbf{92.62} & 73.26 & 63.09 & 91.43 \\
& w/o Prior & 70.52 & 59.69 & 90.44 & 72.09 & 61.30 & 91.44 \\
& DyHist    & \textbf{73.78} & \textbf{63.70} & 91.84 & \textbf{73.73} & \textbf{63.48} & \textbf{92.05} \\
\midrule
\multirow{3}{*}{Nov.}
& Latest    & 30.92 & 19.62 & 53.63 & 32.45 & 21.41 & 55.17 \\
& w/o Prior & 31.83 & 20.73 & 54.94 & 33.69 & 22.28 & 57.42 \\
& DyHist    & \textbf{32.33} & \textbf{20.76} & \textbf{55.43} & \textbf{33.81} & \textbf{22.43} & \textbf{57.49} \\
\bottomrule
\end{tabular}
\end{table}

\subsection{Analysis of Dynamic Historical Prior}
\label{subsec:dynamic_prior_analysis}

\begin{figure}[t]
    \centering
    \includegraphics[width=1\linewidth]{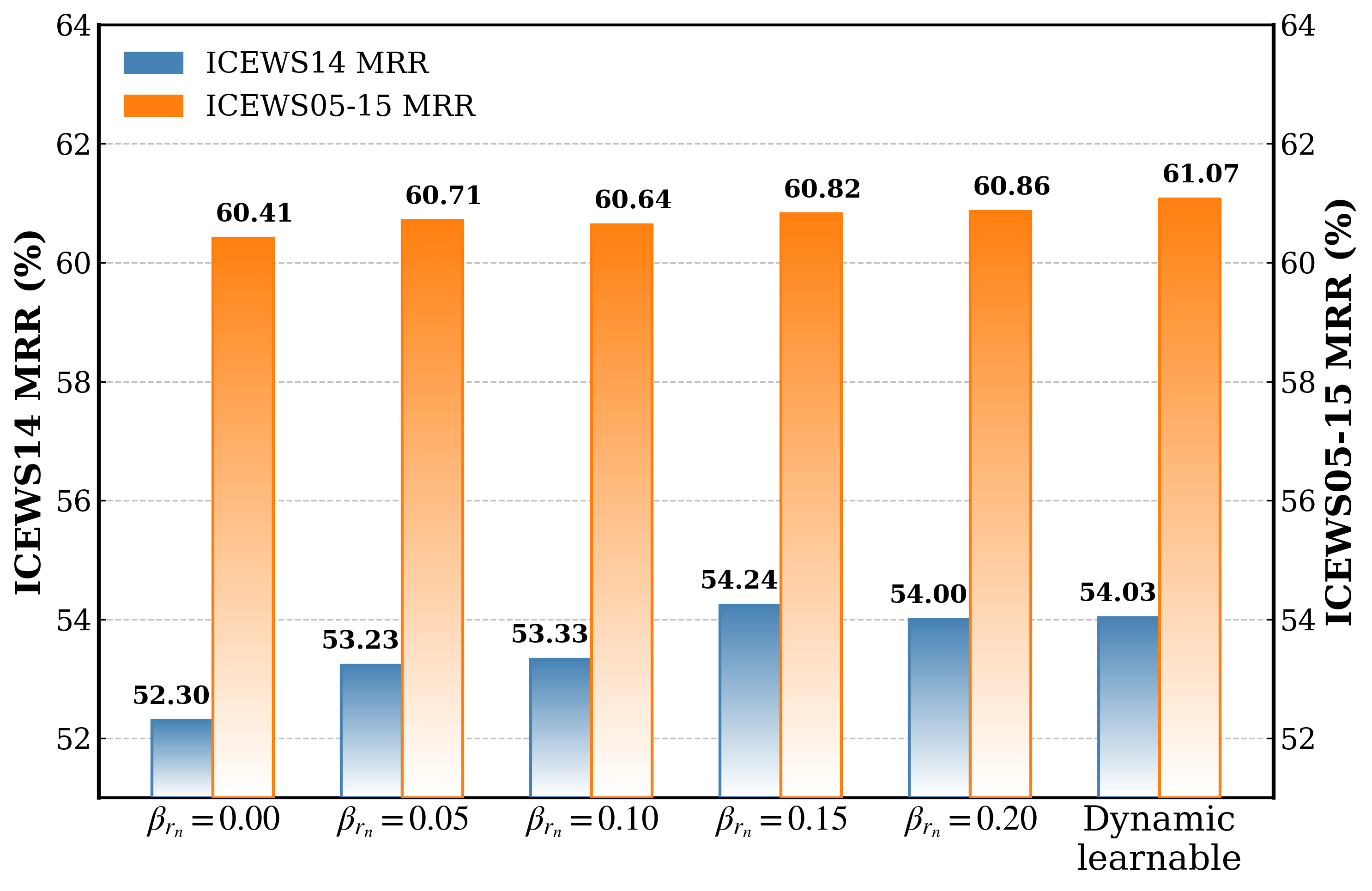}
    \caption{MRR comparison of fixed global and relation-specific learnable recurrence strengths.}
    \label{fig:fixed_vs_dynamic_prior}
\end{figure}

Figure~\ref{fig:fixed_vs_dynamic_prior} compares fixed global prior strengths with the relation-aware learnable prior.
The fixed variants use one shared scalar for all relations, while our variant learns relation-specific strengths $\{\beta_r\}$.
Although a tuned fixed value such as $\beta=0.15$ performs best on ICEWS14, the relation-aware prior achieves the best result on ICEWS05-15, suggesting that learnable relation-specific strengths offer a more adaptive alternative to manually choosing a global coefficient.
We further group relations by history coverage, defined as the proportion of test queries $(s,r,?,t)$ whose subject--relation pair $(s,r)$ appeared before.
This analysis examines whether the learned relation-specific strength $\beta_r$ is related to the recurrence level of different relations.
As shown in Table~\ref{tab:beta_coverage_groups}, $\beta_r$ varies moderately across coverage groups.
On ICEWS14, it increases from low- to high-coverage relations, suggesting that relations with stronger historical recurrence tend to receive larger prior strengths.
On ICEWS05-15, the trend is less monotonic, indicating that coverage alone does not fully determine the learned strength.
Overall, these results suggest that the learned prior provides relation-specific adjustments beyond a uniform recurrence strength.

\begin{table}[h]
\centering
\footnotesize
\setlength{\tabcolsep}{4pt}
\caption{Learned prior strengths under relation-level history coverage groups.}
\label{tab:beta_coverage_groups}
\begin{tabular}{llccc}
\toprule
Dataset & Group & \#Rel. & Avg. Cov. & Avg. $\beta_r$ \\
\midrule
ICEWS14 & Low    & 21 & 49.640 & 0.134 \\
ICEWS14 & Medium & 20 & 76.010 & 0.136 \\
ICEWS14 & High   & 21 & 89.430 & 0.149 \\
\midrule
ICEWS05-15 & Low    & 42 & 55.770 & 0.134 \\
ICEWS05-15 & Medium & 41 & 78.290 & 0.151 \\
ICEWS05-15 & High   & 41 & 91.080 & 0.143 \\
\bottomrule
\end{tabular}
\end{table}

\subsection{Computational Efficiency}
\label{app:efficiency}

\begin{figure}[h]
    \centering
    \includegraphics[width=\linewidth]{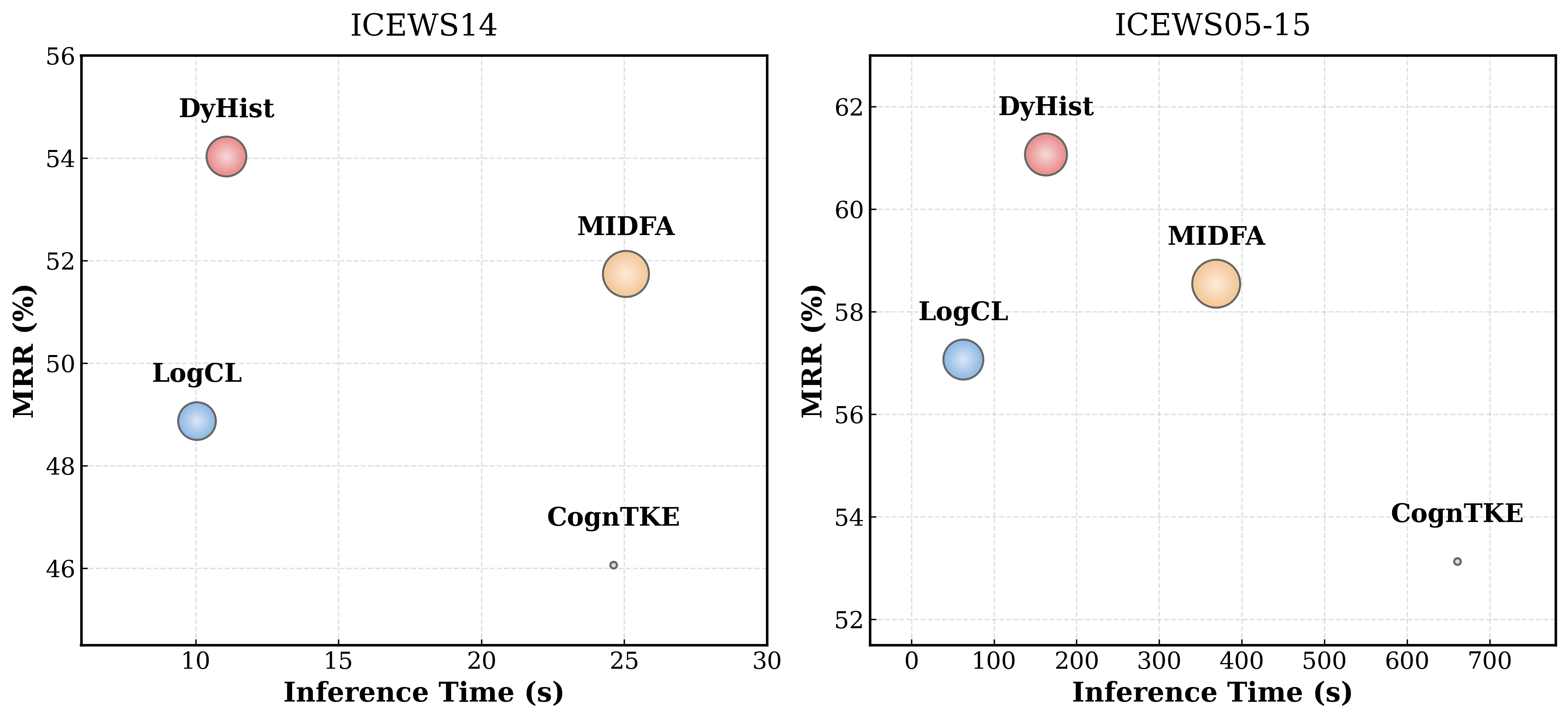}
\caption{Effectiveness--efficiency trade-off on two datasets. Marker size denotes parameter count.}
    \label{fig:efficiency_tradeoff}
\end{figure}

As shown in Figure~\ref{fig:efficiency_tradeoff}, DyHist achieves the highest MRR among the compared models on both datasets.
Its inference time is close to LogCL on ICEWS14 and lower than CognTKE and MIDFA on ICEWS05-15.
Although DyHist uses slightly more parameters than LogCL, it remains smaller than MIDFA.
The additional cost mainly comes from query-aware window selection and candidate-level prior calibration.
Since the historical prior operates at the logit level and does not introduce extra deep decoding layers, the overall inference time remains comparable to efficient baselines.

\section{Conclusion}
\label{sec:conclusion}

In this paper, we proposed DyHist, a query-aware historical dynamics modeling framework for temporal knowledge graph extrapolation.
DyHist models historical evidence from two complementary perspectives: query-relevant local changes and relation-dependent historical recurrence.
It separates recent dynamic evidence from medium-range contextual evidence through adaptive dual-window selection and fast-slow residual modeling, and calibrates recurrence signals with a relation-aware dynamic historical prior.
Experiments on four benchmarks support the usefulness of the proposed historical evidence modeling.
The results suggest that TKG extrapolation requires more than aggregating historical facts: models should also estimate when historical evidence is reliable for the current query.
Historical recurrence provides useful signals, while query-aware local evidence remains important when direct recurrence is weak.

\section{Limitations}
\label{sec:limitations}
One limitation of this work lies in its reliance on four widely used TKG benchmarks, namely ICEWS14, ICEWS05-15, ICEWS18, and GDELT. 
Although standard, these datasets are mainly dominated by political and international event records, with limited diversity in domains such as science, economics, public health, or natural disasters. 
Therefore, the generalizability of DyHist to other types of temporal knowledge graphs remains to be further verified. 
In addition, DyHist explicitly exploits historical recurrence and recent support trends, which may be less effective for sparse and emerging events where direct recurrence is limited.
Moreover, because event-centric TKGs may contain reporting biases, recurrence-based calibration could amplify frequently observed entities or relations if predictions are directly used in sensitive decision-making scenarios.
In future work, we will investigate more effective modeling of such scenarios, where future facts rely more heavily on indirect temporal evidence.
\section{Ethical Considerations}
DyHist is evaluated on public temporal knowledge graph benchmarks and is intended solely for research on temporal reasoning and knowledge graph forecasting.
Since event-centric datasets such as ICEWS and GDELT may contain reporting biases and uneven event coverage, predictions from the model should not be directly used for high-stakes political, security, or humanitarian decision-making without appropriate human oversight and domain-specific validation.
The recurrence-based prior may amplify historically frequent entities or relations, and future real-world applications should therefore include explicit bias assessment and uncertainty-aware evaluation.

\section*{Acknowledgments}
This work was supported by the National Natural Science Foundation of China (U22B2061), the National Key R\&D Program of China (2022YFB4300603), and the Natural Science Foundation of Sichuan, China (2026NSFSC0429).
\bibliography{custom}

\appendix

\section{Notation Summary}
\label{app:notation}

Table~\ref{tab:notation} summarizes the main notation used throughout DyHist.
We include the symbols relevant to historical snapshot modeling, query-aware window selection, local representation aggregation, and relation-aware prior calibration.
For compactness, auxiliary variables that appear only within a single module are explained where they are introduced.

\section{Experimental Details}
\label{app:experimental_details}

\paragraph{Datasets and evaluation metrics.}
\label{app:dataset_statistics}
We evaluate DyHist on four standard TKG extrapolation benchmarks: ICEWS14, ICEWS18, ICEWS05-15, and GDELT.
The ICEWS datasets~\cite{boschee2015icews} are derived from the Integrated Crisis Early Warning System, and GDELT~\cite{leetaru2013gdelt} is collected from the Global Database of Events, Language, and Tone.
Following previous studies~\cite{chen2024local}, we use chronological splits and report MRR and Hits@$\{1,3,10\}$ under the time-aware filtered setting.
Table~\ref{tab:dataset_statistics} summarizes the dataset statistics.
Compared with ICEWS datasets, GDELT has a much finer temporal granularity, with a time interval of 15 minutes.
ICEWS05-15 contains the largest number of entities and timestamps among the ICEWS datasets, making it suitable for evaluating long-range temporal modeling.
The unseen-event ratios show that all datasets contain a considerable proportion of test facts that do not directly appear in the training set, which provides a diagnostic setting beyond directly repeated facts.
\begin{table}[h]
\centering
\footnotesize
\setlength{\tabcolsep}{4pt}
\renewcommand{\arraystretch}{0.95}
\caption{Main notation used in DyHist.}
\label{tab:notation}
\begin{tabular}{ll}
\toprule
Symbol & Description \\
\midrule
$(s_n,r_n,?,t)$ & Future query at timestamp $t$ \\
$\mathcal{G}_{\tau}$ & Temporal snapshot at timestamp $\tau$ \\
$\mathcal{H}_t^{\mathrm{loc}}$ & Local history sequence before $t$ \\
$\mathcal{G}_t^{\mathrm{glob}}$ & Sampled global historical subgraph \\
$\mathcal{I}_t^{\mathrm{loc}}$ & Local candidate snapshot indices \\
$\mathcal{I}_t^{R}$ & Selected recent-window indices \\
$\mathcal{I}_t^{M}$ & Selected medium-window indices \\
$\mathcal{I}_t^{S}$ & Short-term subset of the recent window \\
$\mathcal{I}_t^{\mathrm{rem}}$ & Remaining local indices outside $\mathcal{I}_t^{R}$ \\
\midrule
$\mathbf{e}_{s_n}^{0}$ & Learnable base embedding of subject $s_n$ \\
$\mathbf{e}_{r_n}^{0}$ & Learnable base embedding of relation $r_n$ \\
$\mathbf{e}_{r_n,t}$ & Time-aware relation representation \\
$\mathbf{q}_n$ & Query representation \\
$\mathbf{H}_{l}^{\tau}$ & Local entity-state matrix at timestamp $\tau$ \\
$\mathbf{H}_{g}^{t}$ & Global historical representation at query time \\
$\mathbf{d}_{\tau}$ & Snapshot descriptor of $\mathcal{G}_{\tau}$ \\
\midrule
$\mathbf{h}_{n}^{R}$ & Aggregated recent-window representation \\
$\mathbf{h}_{n}^{M}$ & Aggregated medium-window representation \\
$\mathbf{h}_{n}^{L}$ & Fused local historical context \\
$\mathbf{h}_{n,\mathrm{short}}^{t}$ & Short-term local representation \\
$\Delta_n^t$ & Fast-slow residual deviation \\
$\mathbf{h}_{l,s_n}^{t}$ & Final local subject representation \\
$\mathbf{h}_{s_n,t}^{\mathrm{dec}}$ & Subject-side decoding representation \\
\midrule
$\mathbf{z}_t^{(n)}$ & Neural logits over candidate objects \\
$c_t^{(n)}(o)$ & Time-decayed support of candidate $o$ \\
$\eta_t^{(n)}(o)$ & Recent support trend of candidate $o$ \\
$\beta_{r_n}$ & Relation-specific prior strength \\
$b_t^{(n)}(o)$ & Relation-aware historical bias \\
$\hat{z}_t^{(n)}(o)$ & Calibrated logit of candidate $o$ \\
$\bar{b}_t$ & Reward baseline for policy optimization \\
\bottomrule
\end{tabular}
\end{table}
\paragraph{Additional implementation details.}
We use the same hyperparameter settings across all datasets unless otherwise specified.
The learning rate is set to 0.001, the weight decay is $1\times10^{-5}$, and gradient clipping is applied with a maximum norm of 1.0.
The RGCN encoder uses two layers with 100 bases.
The local history length is set to 10, with both recent and medium window sizes set to 3, and the fast window size set to 2.
For window selection, the policy temperature is set to 0.7, and the entropy weight is set to 0.02.
The dynamic historical prior is applied at the logit level, with initial prior strength 0.11, temporal decay factor $\gamma=0.15$, and trend weight $\rho=0.6$.
The local--global fusion weight $\lambda$ is set to 0.95.
Unless otherwise specified, the reported experiments use a fixed random seed of 3407.

\paragraph{Training and inference.}
Training is conducted at the timestamp level, where each batch contains facts from one timestamp.
We train DyHist and its variants using the same chronological training protocol and select the checkpoint with the best validation MRR for final testing. For window selection, we sample from the learned policy distributions during training, while using top-ranked snapshots during inference to ensure stable predictions.
Unless otherwise specified, all ablation and analysis experiments keep the same backbone, optimizer, and training settings, and only vary the target module or hyperparameter under study.

\section{Baseline Descriptions}
\label{app:baseline_descriptions}

We briefly describe the baseline models used in our experiments.
The compared methods cover static knowledge graph reasoning, temporal interpolation, and temporal extrapolation models.

\paragraph{Static baselines.}
\textbf{DistMult}~\citep{yang2015embedding} uses a bilinear scoring function with diagonal relation matrices to model pairwise interactions between subject and object embeddings.
\textbf{ConvE}~\citep{dettmers2018convolutional} applies two-dimensional convolution over reshaped entity and relation embeddings to capture richer feature interactions for link prediction.
\textbf{RotatE}~\citep{sun2018rotate} represents relations as rotations in the complex embedding space, enabling the model to capture relational patterns such as symmetry, antisymmetry, inversion, and composition.

\paragraph{Interpolation baselines.}
\textbf{TTransE}~\citep{garcia2018learning} extends TransE by incorporating temporal embeddings and models time-dependent facts through translation operations.
\textbf{TA-DistMult}~\citep{garcia2018learning} introduces time-aware relation representations into the DistMult scoring function, allowing relation parameters to vary across temporal contexts.
\textbf{DE-SimplE}~\citep{goel2020diachronic} uses diachronic embeddings that represent entity embeddings as functions of time, thereby modeling temporal changes in entity representations.

\paragraph{Extrapolation baselines.}
\textbf{RE-NET}~\citep{jin2020recurrent} models historical event sequences with recurrent neural networks and aggregates concurrent events for future link prediction.
\textbf{RE-GCN}~\citep{li2021temporal} combines relational graph convolution with recurrent evolution to update entity representations over temporal snapshots.
\textbf{CyGNet}~\citep{zhu2021learning} models temporal recurrence through a copy-generation framework that balances historical copying and candidate generation.
\textbf{TiRGN}~\citep{li2022tirgn} encodes historical snapshots with recurrent graph neural networks to capture temporal relational evolution for extrapolation.
\textbf{HisMatch}~\citep{li2022hismatch} matches current queries with historical facts and exploits historical consistency to improve future entity prediction.
\textbf{RETIA}~\citep{liu2023retia} constructs temporal interaction-aware representations by modeling relational evolution and aggregating historical neighborhood information.
\textbf{CENET}~\citep{xu2023temporal} exploits contrastive event representation learning and historical candidate information to enhance temporal extrapolation.
\textbf{CRAFT}~\citep{zhang2024modeling} models coarse- and fine-grained historical contexts to capture multi-level temporal dependencies for future fact prediction.
\textbf{THCN}~\citep{chen2024thcn} introduces temporal causal convolutional networks to model the influence of historical and concurrent facts in temporal knowledge graphs.
\textbf{LogCL}~\citep{chen2024local} learns local historical representations with contrastive learning, aiming to improve the discrimination of query-relevant temporal contexts.
\textbf{CognTKE}~\citep{chen2025cogntke} incorporates query-related cognitive subgraphs and symbolic evidence to improve interpretable temporal extrapolation.
\textbf{NADEx}~\citep{gan26negative} introduces a negative-aware diffusion framework that reconstructs query objects with a Transformer denoiser conditioned on temporal-relational context and regularized by negative prototypes.
\textbf{MIDFA}~\citep{wang2026temporal} models multidimensional information interaction and dynamic frequency awareness, jointly capturing semantic, temporal, structural, and time-sensitive frequency signals for TKG reasoning.

\begin{table}[t]
\centering
\footnotesize
\setlength{\tabcolsep}{1pt}
\caption{Statistics of the datasets.}
\label{tab:dataset_statistics}
\begin{tabular}{lrrrr}
\toprule
Dataset & ICEWS14 & ICEWS18 & ICEWS05-15 & GDELT \\
\midrule
Training Facts   & 74,845  & 373,018 & 368,868   & 1,734,399 \\
Validation Facts & 8,514   & 45,995  & 46,302    & 238,765 \\
Test Facts       & 7,371   & 49,545  & 46,159    & 305,241 \\
Entities         & 6,869   & 10,094  & 23,033    & 7,691 \\
Relations        & 230     & 256     & 251       & 240 \\
Time Interval    & 1 day   & 1 day   & 1 day     & 15 mins \\
Timestamps       & 365     & 365     & 4,017     & 2,975 \\
Unseen Events    & 58.43\% & 55.69\% & 39.82\%   & 43.72\% \\
\bottomrule
\end{tabular}
\end{table}

\section{Details of Query-aware Local Dynamics Modeling}
\label{app:local_dynamics_details}
This section provides additional details of the query-aware local dynamics modeling module, including snapshot descriptor construction, dual-window selection, and policy optimization.

\subsection{Snapshot Descriptor Construction}
\label{app:descriptor_details}

For each target timestamp $t$, DyHist uses the local history
$\mathcal{H}_t^{\mathrm{loc}}=\{\mathcal{G}_{t-L},\ldots,\mathcal{G}_{t-1}\}$
as the candidate span, where $L=10$ in our main setting.
After encoding these snapshots with the local recurrent RGCN encoder, each candidate snapshot
$\mathcal{G}_{\tau}$ is represented by the local entity-state matrix
$\mathbf{H}_{l}^{\tau}\in\mathbb{R}^{|\mathcal{E}|\times d}$.
Let $\mathcal{B}_t=\{(s_i,r_i,o_i,t)\}_{i=1}^{N_t}$ denote the timestamp-level batch.
For each query $(s_i,r_i,?,t)$, we construct its query representation
$\mathbf{q}_i$ from the subject and relation embeddings.
DyHist then builds compact snapshot descriptors to guide query-aware local window selection.

\paragraph{Query-snapshot relevance.}
The first feature measures how relevant a candidate snapshot is to the current timestamp-level queries.
For each snapshot $\mathcal{G}_{\tau}$, we compute the average cosine similarity between the subject-side local states and the corresponding query representations:
\begin{equation}
\tilde{u}_{\tau}
=
\frac{1}{N_t}
\sum_{i=1}^{N_t}
\cos
\left(
\mathbf{H}_{l}^{\tau}[s_i],
\mathbf{q}_i
\right),
\end{equation}
where $\mathbf{H}_{l}^{\tau}[s_i]$ is the local state of subject entity $s_i$ at snapshot $\tau$.
This feature favors snapshots whose subject-side states are semantically aligned with the current query batch.

\paragraph{Temporal lag.}
The second feature encodes the relative temporal distance from the query timestamp:
\begin{equation}
\ell_{\tau}
=
\frac{t-1-\tau}{L-1},
\qquad
\ell_{\tau}\in[0,1].
\end{equation}
Here, $\ell_{\tau}=0$ corresponds to the latest snapshot $\mathcal{G}_{t-1}$, while
$\ell_{\tau}=1$ corresponds to the earliest snapshot $\mathcal{G}_{t-L}$.
This feature helps distinguish near-query evidence from earlier contextual evidence.

\paragraph{Local variation.}
Unlike the query-specific relevance features, the following two features describe snapshot-level structural dynamics by aggregating the local states of all entities.
The third feature captures the state transition between adjacent snapshots. 
For $\tau>t-L$, we compute
\begin{equation}
\tilde{\delta}_{\tau}
=
\frac{1}{|\mathcal{E}|}
\sum_{e\in\mathcal{E}}
\left\|
\mathbf{H}_{l}^{\tau}[e]
-
\mathbf{H}_{l}^{\tau-1}[e]
\right\|_2 .
\end{equation}
For the earliest snapshot, we set $\tilde{\delta}_{t-L}=0$.
A larger $\tilde{\delta}_{\tau}$ indicates a stronger global state change around snapshot $\tau$,
which helps identify snapshots containing dynamic shifts relevant to future prediction.

\paragraph{Latest-state consistency.}
The fourth feature measures how consistent a candidate snapshot is with the latest local state:
\begin{equation}
\tilde{\alpha}_{\tau}
=
\frac{1}{|\mathcal{E}|}
\sum_{e\in\mathcal{E}}
\cos
\left(
\mathbf{H}_{l}^{\tau}[e],
\mathbf{H}_{l}^{t-1}[e]
\right).
\end{equation}
This feature complements the local variation feature by helping the policy avoid selecting snapshots that show large changes but are no longer consistent with the latest local context.

\paragraph{Relation-context alignment.}
In addition to subject-side query relevance, DyHist computes a relation-context feature to measure whether a candidate snapshot is aligned with the relation context of the current batch.
Let $\mathbf{c}_i=W_c[\mathbf{q}_i\Vert\mathbf{e}_{r_i}^{0}]$ denote the relation-aware query context representation,
where $W_c$ is a learnable projection matrix.
We summarize snapshot $\mathcal{G}_{\tau}$ by averaging the local states of all entities:
\begin{equation}
\bar{\mathbf{h}}_{\tau}
=
\frac{1}{|\mathcal{E}|}
\sum_{e\in\mathcal{E}}
\mathbf{H}_{l}^{\tau}[e].
\end{equation}
The alignment score is computed as
\begin{equation}
\tilde{m}_{\tau}
=
\frac{1}{N_t}
\sum_{i=1}^{N_t}
\cos
\left(
\bar{\mathbf{h}}_{\tau},
\mathbf{c}_i
\right).
\end{equation}
After normalization, we use $m_{\tau}$ and its recentness-adjusted form $m_{\tau}(1-\ell_{\tau})$ as descriptor features.

\paragraph{Feature normalization.}
Since the descriptor features have different numerical scales, we standardize all non-lag features over the candidate span:
\begin{equation}
x_{\tau}
=
\frac{\tilde{x}_{\tau}-\mu_x}{\sigma_x+\epsilon},
\
\tilde{x}_{\tau}
\in
\{
\tilde{u}_{\tau},
\tilde{\delta}_{\tau},
\tilde{\alpha}_{\tau},
\tilde{m}_{\tau}
\},
\end{equation}
where $\mu_x$ and $\sigma_x$ are computed over
$\{\mathcal{G}_{t-L},\ldots,\mathcal{G}_{t-1}\}$. The normalized features are denoted as
$u_{\tau}$, $\delta_{\tau}$, $\alpha_{\tau}$, and $m_{\tau}$, while the temporal lag $\ell_{\tau}$ is already normalized into $[0,1]$.
The final snapshot descriptor is written as
\begin{equation}
\mathbf{d}_{\tau}
=
[
u_{\tau},
\ell_{\tau},
\delta_{\tau},
\alpha_{\tau},
m_{\tau},
m_{\tau}(1-\ell_{\tau})
].
\end{equation}
For implementation, we split $\mathbf{d}_{\tau}$ into two parts:
$\mathbf{d}_{\tau}=[\boldsymbol{\phi}_{\tau}\|\boldsymbol{\psi}_{\tau}]$,
where $\boldsymbol{\phi}_{\tau}=[u_{\tau},\ell_{\tau},\delta_{\tau},\alpha_{\tau}]$
summarizes query relevance, temporal distance, local variation, and latest-state consistency,
and $\boldsymbol{\psi}_{\tau}=[m_{\tau},m_{\tau}(1-\ell_{\tau})]$
encodes relation-context alignment and its recentness-adjusted form.

\paragraph{Recent and medium window scoring.}
Let $\mathcal{I}_t^{\mathrm{loc}}=\{t-L,\ldots,t-1\}$ denote the candidate snapshot indices.
For each branch $X\in\{R,M\}$, DyHist computes selection logits over snapshot descriptors:
\begin{equation}
\zeta_{\tau}^{X}
=
f_X([\boldsymbol{\phi}_{\tau}\Vert\boldsymbol{\psi}_{\tau}]).
\end{equation}
The policy distribution is
\begin{equation}
\pi_X(\tau\mid\mathcal{D}_t)
=
\mathrm{softmax}_{\tau\in\mathcal{C}_t^X}
(\zeta_{\tau}^{X}),
\end{equation}
where $\mathcal{C}_t^R=\mathcal{I}_t^{\mathrm{loc}}$ and
$\mathcal{C}_t^M=\mathcal{I}_t^{\mathrm{loc}}\setminus\mathcal{I}_t^R$.
Thus, the medium window is selected from snapshots not used by the recent window.

\subsection{Query-guided Temporal Aggregation}
\label{app:aggregation_details}

After selecting the recent and medium windows, DyHist aggregates the encoded subject states within each window.
For query $(s_n,r_n,?,t)$ and snapshot $\mathcal{G}_{\tau}$, we first compute a query-guided historical representation:
\begin{equation}
\mathbf{a}_{n}^{\tau}
=
\phi_a
\left(
[
\mathbf{H}_{l}^{\tau}[s_n]
\Vert
\mathbf{q}_n
]
\right),
\end{equation}
where $\phi_a(\cdot)$ is a projection network.
For each window $X\in\{R,M\}$, the aggregation weight is computed as
\begin{equation}
\omega_{n,\tau}^{X}
=
\mathrm{softmax}_{\tau\in\mathcal{I}_t^X}
\left(
g_X([\mathbf{a}_{n}^{\tau}\Vert \mathbf{q}_n\Vert \mathbf{d}_{\tau}])
\right),
\end{equation}
where $g_X(\cdot)$ is a window-specific scoring function.
The window representation is then obtained by
\begin{equation}
\mathbf{h}_{n}^{X}
=
\sum_{\tau\in\mathcal{I}_t^X}
\omega_{n,\tau}^{X}
\mathbf{a}_{n}^{\tau}.
\end{equation}
This implements the query-guided temporal aggregator $\mathrm{Agg}_X(\cdot)$ used in the main text.
\paragraph{Short-term aggregation.}
For the short-term subset $\mathcal{I}_t^S\subseteq\mathcal{I}_t^R$, which contains the most recent snapshots in the recent window, DyHist uses the same aggregation form:
\begin{equation}
\mathbf{h}_{n,\mathrm{short}}^{t}
=
\sum_{\tau\in\mathcal{I}_t^S}
\omega_{n,\tau}^{S}
\mathbf{a}_{n}^{\tau},
\end{equation}
where $\omega_{n,\tau}^{S}$ is computed by the recent-window scoring function $g_R(\cdot)$ restricted to $\mathcal{I}_t^S$.
This short-term representation serves as the fast evidence used to compute the residual deviation from the broader recent--medium historical context.

\subsection{Window Selection Loss}
\label{app:pg_objective}

DyHist optimizes the recent and medium window policies with prediction-oriented feedback and temporal-role regularization.
For each timestamp batch $t$, the policy selects a recent window $\mathcal{I}_t^{R}$ and a medium window $\mathcal{I}_t^{M}$ from the local candidate history.
Let
\begin{equation}
a_t=(\mathcal{I}_t^{R},\mathcal{I}_t^{M})
\end{equation}
denote the joint window-selection action, and let $\mathcal{D}_t$ denote the descriptor set of candidate snapshots.

The reward combines prediction feedback and temporal-role separation:
\begin{equation}
\mathcal{R}_t
=
\lambda_m\mathcal{R}_{\mathrm{mar}}
+
\lambda_n\mathcal{R}_{\mathrm{nll}}
+
\mathcal{R}_{\mathrm{sep}}.
\end{equation}
The margin reward measures whether the selected windows help separate the ground-truth entity from the strongest negative candidate:
\begin{equation}
\mathcal{R}_{\mathrm{mar}}
=
\frac{1}{N_t}
\sum_{n=1}^{N_t}
\left(
\hat{z}_t^{(n)}(o_n)
-
\max_{o\neq o_n}
\hat{z}_t^{(n)}(o)
\right),
\end{equation}
where $\hat{z}_t^{(n)}(o)$ denotes the calibrated logit of candidate object $o$ for query $(s_n,r_n,?,t)$.
The NLL reward is defined as
\begin{equation}
\mathcal{R}_{\mathrm{nll}}
=
-\mathcal{L}_{\mathrm{ent}},
\end{equation}
so lower entity prediction loss leads to a larger reward.
The separation term encourages the two windows to capture different temporal roles:
\begin{equation}
\mathcal{R}_{\mathrm{sep}}
=
\lambda_{\mathrm{div}}\mathcal{D}_{\phi}
+
\lambda_{\mathrm{lag}}\mathcal{D}_{\ell}.
\end{equation}
Here, $\mathcal{D}_{\phi}$ measures the descriptor-level difference between the selected recent and medium windows:
\begin{equation}
\mathcal{D}_{\phi}
=
\left\|
\frac{1}{|\mathcal{I}_t^R|}
\sum_{\tau\in\mathcal{I}_t^R}
\boldsymbol{\phi}_{\tau}
-
\frac{1}{|\mathcal{I}_t^M|}
\sum_{\tau\in\mathcal{I}_t^M}
\boldsymbol{\phi}_{\tau}
\right\|_2,
\end{equation}
and $\mathcal{D}_{\ell}$ measures their temporal lag separation:
\begin{equation}
\mathcal{D}_{\ell}
=
\frac{1}{|\mathcal{I}_t^M|}
\sum_{\tau\in\mathcal{I}_t^M}
\ell_{\tau}
-
\frac{1}{|\mathcal{I}_t^R|}
\sum_{\tau\in\mathcal{I}_t^R}
\ell_{\tau}.
\end{equation}
Thus, the separation term encourages the recent window to focus on near-query dynamic evidence and the medium window to preserve earlier contextual information.
The window-selection policy is optimized with a policy-gradient objective:
\begin{equation}
\mathcal{L}_{\mathrm{pg}}
=
-
(\mathcal{R}_t-\bar{b}_t)
\log \pi(a_t\mid \mathcal{D}_t)
-
\lambda_{\mathrm{H}}\mathcal{H}(\pi),
\end{equation}
where $\bar{b}_t$ denotes the moving average of historical rewards and $\mathcal{H}(\pi)$ is an entropy regularizer.
The term $\mathcal{R}_t-\bar{b}_t$ measures whether the current window selection performs better than the average reward level.

\section{Repeated and Novel Queries}
\label{app:repeated_novel}

We further examine how DyHist behaves on repeated and less repetitive queries.
Following the occurrence of the ground-truth object in history, we split test queries into repeated and novel subsets.
A query is considered repeated if its ground-truth object has appeared with the same subject--relation pair in historical snapshots; otherwise, it is considered novel.
The split is query-level and fixed across all compared models. Table~\ref{tab:repeat_novel_stats} summarizes the subset statistics.
ICEWS14 contains 7,371 test queries, among which 3,860 are repeated and 3,511 are novel.
ICEWS05-15 contains 46,159 test queries, among which 31,567 are repeated and 14,592 are novel.
The repeated ratio is 52.37\% on ICEWS14 and 68.39\% on ICEWS05-15, indicating that ICEWS05-15 contains stronger historical recurrence.
This makes ICEWS05-15 suitable for evaluating whether the dynamic historical prior can exploit recurrence patterns.
\begin{table}[t]
  \centering
  \footnotesize
  \setlength{\tabcolsep}{1pt}
  \caption{Statistics of the repeated/novel split.}
  \label{tab:repeat_novel_stats}
  \begin{tabular}{lrrrrr}
  \toprule
  Dataset & \#Test & \#Repeated & \#Novel & Rep. (\%) & Nov. (\%) \\
  \midrule
  ICEWS14    & 7,371  & 3,860  & 3,511  & 52.37 & 47.63 \\
  ICEWS05-15 & 46,159 & 31,567 & 14,592 & 68.39 & 31.61 \\
  \bottomrule
  \end{tabular}
\end{table}

 \section{Details of Dynamic Historical Prior Calibration}
  \label{app:dynamic_prior_details}

  This section provides additional details of the dynamic historical prior used in the decoder.
  Different from representation-level historical fusion, DyHist calibrates candidate logits with a query-conditioned historical prior.
  For a query $(s,r,?,t)$, the historical prior is constructed by retrieving historical facts that share the same subject--relation pair $(s,r)$. Thus, a candidate object receives historical support only if the same subject--relation pair $(s_n,r_n)$ has linked to it in previous snapshots.
\paragraph{Context-aware relation representation.}
We maintain a learnable base relation embedding $\mathbf{e}_{r}^{0}$ for each relation.
At each historical snapshot $\mathcal{G}_{\tau}$, DyHist aggregates the states of
entities associated with relation $r$ to obtain a relation-specific context
$\mathbf{c}_{r}^{\tau}$, which is fused with $\mathbf{e}_{r}^{0}$ through a sigmoid gate:
\begin{equation}
\mathbf{x}_{r}^{\tau}
=
\mathbf{e}_{r}^{0}
+
\mathbf{c}_{r}^{\tau},
\qquad
\mathbf{g}_{r}^{\tau}
=
\sigma(\mathbf{x}_{r}^{\tau}W_g+\mathbf{b}_g),
\end{equation}
where $W_g$ and $\mathbf{b}_g$ denote the shared learnable weight matrix and bias vector of the gating function, respectively.
\begin{equation}
\mathbf{e}_{r}^{\tau}
=
\mathbf{g}_{r}^{\tau}\odot\mathbf{x}_{r}^{\tau}
+
(1-\mathbf{g}_{r}^{\tau})\odot\mathbf{e}_{r}^{0}.
\end{equation}
The final representation is used as $\mathbf{e}_{r,t}$ in the decoder.

\subsection{Query-conditioned Historical Support}
\label{app:historical_support}

Given the query $(s_n,r_n,?,t)$ and a candidate object $o$, DyHist first computes its decayed historical support under the current subject--relation pair:
\begin{equation}
c_t^{(n)}(o)
=
\sum_{\tau<t}
e^{-\gamma(t-\tau)}
\mathbb{I}\left[(s_n,r_n,o)\in\mathcal{G}_{\tau}\right],
\end{equation}
where $\gamma$ is the temporal decay coefficient and $\mathbb{I}[\cdot]$ is the indicator function.
This term assigns larger weights to more recent historical occurrences while still preserving long-range recurrence evidence.

To capture whether a candidate is becoming more relevant near the query timestamp, DyHist further introduces a recent trend term.
Let $\mathcal{I}_{t}^{\mathrm{loc}}=\{t-L,\ldots,t-1\}$ denote the local candidate snapshot indices, and let
$\mathcal{I}_{t}^{\mathrm{rem}}=\mathcal{I}_{t}^{\mathrm{loc}}\setminus \mathcal{I}_{t}^{R}$
denote the remaining local indices.
We compute the occurrence counts in the recent window and the remaining local history as
\begin{equation}
c_{t,\mathrm{rec}}^{(n)}(o)
=
\sum_{\tau\in\mathcal{I}_{t}^{R}}
\mathbb{I}\left[(s_n,r_n,o)\in\mathcal{G}_{\tau}\right].
\end{equation}
\begin{equation}
c_{t,\mathrm{rem}}^{(n)}(o)
=
\sum_{\tau\in\mathcal{I}_{t}^{\mathrm{rem}}}
\mathbb{I}\left[(s_n,r_n,o)\in\mathcal{G}_{\tau}\right].
\end{equation}
The recent trend score is then defined by comparing their normalized occurrence rates:
\begin{equation}
\eta_t^{(n)}(o)
=
\mathrm{ReLU}
\left(
\frac{c_{t,\mathrm{rec}}^{(n)}(o)}{|\mathcal{I}_{t}^{R}|}
-
\frac{c_{t,\mathrm{rem}}^{(n)}(o)}{|\mathcal{I}_{t}^{\mathrm{rem}}|}
\right).
\end{equation}
The trend term $\eta_t^{(n)}(o)$ emphasizes candidates whose support increases near the query time, rather than simply favoring candidates that are frequent in distant history.
The relation-aware historical bias is then computed as
\begin{equation}
b_t^{(n)}(o)
=
\beta_{r_n}
\left[
\log\left(1+c_t^{(n)}(o)\right)
+
\rho \eta_t^{(n)}(o)
\right],
\end{equation}
where $\beta_{r_n}$ controls the relation-specific recurrence strength and $\rho$ controls the contribution of the recent trend term.
The logarithmic transformation prevents highly frequent candidates from dominating the neural decoder.

\subsection{Relation-aware Prior Strength}
\label{app:relation_aware_prior_strength}

The relation-aware historical bias is added to the neural decoder logit as a candidate-level calibration term:
\begin{equation}
\hat{z}_t^{(n)}(o)
=
z_t^{(n)}(o)
+
b_t^{(n)}(o),
\end{equation}
where $z_t^{(n)}(o)$ is the neural decoder logit and $\hat{z}_t^{(n)}(o)$ is the calibrated logit.
The strength of the historical bias is controlled by the relation-specific parameter $\beta_{r_n}$.
Compared with a fixed global strength shared by all relations, relation-aware strength allows DyHist to assign stronger recurrence calibration to repetitive relations and weaker calibration to event-driven relations.
This design allows historical recurrence to be weighted differently across relation types.

\section{Case Study}
\label{app:case_study}

We provide case studies on ICEWS14 to illustrate how DyHist uses local historical evidence and dynamic historical prior calibration.
Different from quantitative ablation results, the case studies focus on individual test queries and show how the rank of the ground-truth entity changes under different model variants or scoring stages.
We use filtered ranks throughout this section.

\subsection{Counterfactual Ranking Analysis}
\label{app:case_counterfactual}

We first compare DyHist with several degraded variants on the same query.
For the query $(\textit{Amnesty International}, \textit{Make statement}, ?, t)$, the ground-truth object is \textit{Israeli Defense Forces}.
This query is challenging because several distracting candidates, such as police-related entities and country entities, receive high neural decoding scores.

\begin{table}[t]
\centering
\footnotesize
\setlength{\tabcolsep}{1pt}
\caption{Counterfactual case study on ICEWS14. The same query is evaluated under different model variants.}
\label{tab:case_counterfactual}

\begin{tabularx}{\columnwidth}{l c >{\raggedright\arraybackslash}X}
\toprule
Model Variant & GT Rank & \multicolumn{1}{c}{Top-3 Predictions} \\
\midrule
w/o Prior
& 10
& Police (Hong Kong), Iran, Police (Canada) \\

w/o QADW
& 4
& Police (Hong Kong), Government (Nigeria), Iran \\

DyHist
& \textbf{1}
& \textbf{Israeli Defense Forces}, Boko Haram, Iran \\
\bottomrule
\end{tabularx}

\end{table}

\begin{table*}[t]
\centering
\footnotesize
\setlength{\tabcolsep}{5pt}
\renewcommand{\arraystretch}{1.08}
\caption{Successful-promotion case on ICEWS14. The dynamic prior improves the ground-truth rank from 19 to 1.}
\label{tab:case_detail_amnesty}
\begin{tabular}{p{3.2cm}p{5.8cm}p{5.8cm}}
\toprule
Query & Before Prior & After Prior \\
\midrule
\textbf{Query:}

$(\textit{Amnesty International},$
$\textit{Make statement}, ?, t)$

\textbf{Answer:}

\gt{\textit{Israeli Defense Forces}}
&
GT rank: 19

Top-5:

Boko Haram,

Iran,

France,

Thailand,

China
&
GT rank: \textbf{1}

Top-5:

\gt{\textit{Israeli Defense Forces}},

Boko Haram,

Iran,

France,

Thailand
\\
\midrule
\multicolumn{3}{l}{\textbf{Selected historical evidence from dual windows}} \\
\midrule
\textbf{Recent window}
&
\multicolumn{2}{p{11.6cm}}{
$t-5$: Amnesty International $\xrightarrow{\textit{Make statement}}$ Criminal (Philippines);
$t-4$: Barack Obama $\xrightarrow{\textit{Make statement}}$ Foreign Affairs (Trinidad and Tobago);
John Kerry $\xrightarrow{\textit{Make statement}}$ Iran.
}
\\
\textbf{Medium window}
&
\multicolumn{2}{p{11.6cm}}{
$t-8$: Amnesty International $\xrightarrow{\textit{Make statement}}$ Police (Philippines);
Media (Palestinian Territory, Occupied) $\xrightarrow{\textit{Make statement}}$ \keyev{Israeli Defense Forces}.
$t-7$: John Kerry $\xrightarrow{\textit{Make statement}}$ Military (Russia).
}
\\
\bottomrule
\end{tabular}
\end{table*}

\begin{table*}[t]
\centering
\footnotesize
\setlength{\tabcolsep}{5pt}
\renewcommand{\arraystretch}{1.08}
\caption{Partial-promotion case on ICEWS14. The dynamic prior improves the ground-truth rank from 18 to 3 without forcing it to rank first.}
\label{tab:case_detail_chuck}
\begin{tabular}{p{3.2cm}p{5.8cm}p{5.8cm}}
\toprule
Query & Before Prior & After Prior \\
\midrule
\textbf{Query:}

$(\textit{Chuck Hagel},$
$\textit{Consult}, ?, t)$

\textbf{Answer:}

\gt{\textit{Iraq}}
&
GT rank: 18

Top-5:

Jean-Yves Le Drian,

UN Security Council,

Tamar Beruchashvili,

Abdelkader Messahel,

Alexis Tsipras
&
GT rank: \textbf{3}

Top-5:

Jean-Yves Le Drian,

UN Security Council,

\gt{\textit{Iraq}},

Ashraf Ghani Ahmadzai,

Tamar Beruchashvili
\\
\midrule
\multicolumn{3}{l}{\textbf{Selected historical evidence from dual windows}} \\
\midrule
\textbf{Recent window}
&
\multicolumn{2}{p{11.6cm}}{
$t-4$: Iran $\xrightarrow{\textit{Make statement}}$ \keyev{Iraq};
Terrorist (Islamic State of Iraq and the Levant) $\xrightarrow{\textit{Threaten}}$ \keyev{Iraq}.
$t-5$: Iran $\xrightarrow{\textit{Express intent to meet or negotiate}}$ \keyev{Iraq}.
}
\\
\textbf{Medium window}
&
\multicolumn{2}{p{11.6cm}}{
$t-10$: North Atlantic Treaty Organization $\xrightarrow{\textit{Consult}}$ \keyev{Iraq}.
$t-7$: John Kerry $\xrightarrow{\textit{Make statement}}$ \keyev{Iraq};
Haider Al-Abadi $\xrightarrow{\textit{Make statement}}$ \keyev{Iraq}.
}
\\
\bottomrule
\end{tabular}
\end{table*}

As shown in Table~\ref{tab:case_counterfactual}, removing the dynamic historical prior lowers the rank of the ground-truth entity from 1 to 10.
This indicates that candidate-level historical calibration is important for promoting historically plausible answers.
Removing QADW replaces explicit window selection and fast-slow residual injection with query-aware aggregation over the whole local history.
This hurts the rank, suggesting the benefit of adaptive local evidence modeling.
DyHist ranks the correct entity first, showing that local evidence modeling and dynamic prior calibration provide complementary benefits.

\subsection{Detailed Evidence Utilization}
\label{app:case_evidence}

We further inspect the selected historical evidence and the prediction change before and after applying the relation-aware historical prior.
This analysis explains why the dynamic prior improves the final ranking and whether the selected local history contains meaningful contextual signals.

Table~\ref{tab:case_detail_amnesty} gives a detailed view of the same query used in the counterfactual study.
Before prior calibration, \textit{Israeli Defense Forces} is ranked only 19th, while several globally frequent entities receive higher scores.
The selected evidence shows complementary roles of the two windows: the recent window captures statement-related local context around the query, whereas the medium window retrieves answer-related evidence involving \textit{Israeli Defense Forces}.
After applying the dynamic historical prior, the ground-truth entity is promoted to rank 1.
This suggests that DyHist benefits from combining query-aware local evidence selection with relation-aware historical calibration, rather than relying on either local context or recurrence signals alone.

\subsection{Additional Case: Partial Promotion}
\label{app:case_partial_promotion}

The previous case shows a successful promotion to rank 1.
To avoid relying only on a single highly favorable example, we provide another case where the dynamic prior substantially improves the ground-truth rank but does not make it the top-1 prediction.
This case shows that the prior acts as a calibration signal rather than a deterministic memorization rule.

As shown in Table~\ref{tab:case_detail_chuck}, before prior calibration,
\textit{Iraq} is ranked 18th.
The selected recent evidence contains Iraq-centered regional signals, such as statements, threats, and negotiation intents involving \textit{Iraq}.
These facts indicate that \textit{Iraq} is highly active in the near-query context.
The medium evidence further provides earlier contextual support, including a consultation-related event involving \textit{Iraq}.
This is consistent with the query relation \textit{Consult} and helps connect the candidate entity to the current prediction.
After applying the dynamic historical prior, the rank of \textit{Iraq} improves from 18 to 3.
However, DyHist does not force \textit{Iraq} to become the top-1 prediction, even though it receives a positive prior bias.
This behavior suggests that the historical prior complements neural evidence rather than dominating the decoder.
Therefore, the dynamic prior is better interpreted as a calibration mechanism rather than a hard recurrence-matching rule.

\subsection{Discussion}
\label{app:case_discussion}

These cases provide several observations.
First, the selected dual-window evidence contains query-relevant historical facts, suggesting that local selection can provide interpretable support for prediction.
Second, recent and medium windows capture complementary evidence, with recent windows reflecting near-query contexts and medium windows providing earlier semantic clues.
Third, the dynamic historical prior calibrates candidate scores according to recurrence evidence: it can promote strongly supported ground-truth objects to the top rank, while also improving ranks without acting as a hard copy rule.
Together, these examples support the main intuition of DyHist: temporal extrapolation benefits from adaptive local history modeling and dynamic candidate-level historical calibration.

\end{document}